\documentclass[12pt,a4paper]{article}
\usepackage{graphicx}
\usepackage{amssymb}
\usepackage{amsmath}
\usepackage{bm}
\usepackage{color}
\usepackage{theorem}
\usepackage{subfigure}

\usepackage[sort&compress,numbers, merge]{natbib}

\definecolor{Orange}{cmyk}{0,0.61,0.87,0}
\definecolor{JungleGreen}{cmyk}{0.99,0,0.52,0}
\definecolor{OliveGreen}{cmyk}{0.64,0,0.95,0.40}
\definecolor{Brown}{cmyk}{0,0.81,1,0.60}
\definecolor{RoyalBlue}{cmyk}{0.71,0.53,0,0.12}

\usepackage[colorlinks=True, citecolor=blue, linkcolor=blue, urlcolor=black]{hyperref}

\allowdisplaybreaks[1]

\begin{document}

\begin{titlepage}
\begin{center}
\hfill UMN--TH--3909/20\\ 
\hfill IPPP-19-95

\vspace{2.0cm}
{\Large\bf The Axion Mass from  5D Small Instantons}

\vspace{1.0cm}
{\small \bf Tony Gherghetta$^{a}$,
Valentin V. Khoze$^{b}$,
Alex Pomarol$^{c,d}$,
Yuri Shirman$^{e}$}

\vspace{0.5cm}
{\it\footnotesize
${}^a$School of Physics and Astronomy, University of Minnesota, Minneapolis, Minnesota 55455, USA\\
${}^b$Institute for Particle Physics Phenomenology, Department of Physics, 
Durham University, \\Durham, DH1 3LE, United Kingdom\\
${}^c$IFAE and BIST, Universitat Autonoma de Barcelona, 08193 Bellaterra, Barcelona\\
${}^d$Departmento de Fisica, Universitat Autonoma de Barcelona, 08193 Bellaterra, Barcelona\\
${}^e$ Department of Physics and Astronomy, University of California, Irvine, California 92697, USA
}

\vspace{0.5cm}
\abstract
We calculate a new contribution to the axion mass that arises 
from gluons propagating in a 5th dimension at high energies.
By uplifting the 4D instanton solution to five dimensions, the positive frequency modes 
of the Kaluza-Klein states generate a power-law term in the effective action that inversely grows
with the instanton size. This causes 5D small instantons to enhance the axion mass in a way 
that does not spoil the axion solution to the strong CP problem. Moreover this enhancement can be much 
larger than the usual QCD contribution from large instantons, although it requires the 5D gauge theory to be near the non-perturbative limit. Thus our result suggests that the mass range of axions (or axion-like particles), which is important for ongoing experimental searches, can depend sensitively on the UV modification of QCD.

\end{center}
\end{titlepage}
\setcounter{footnote}{0}

\section{Introduction}

The axion is arguably the best motivated new particle beyond the Standard Model (SM). Its existence is required by the Peccei-Quinn mechanism~\cite{Peccei:1977hh}, which is a popular solution to the strong CP problem. In particular, the axion is identified with the Nambu-Goldstone boson~\cite{Weinberg:1977ma,Wilczek:1977pj} that arises from a spontaneously broken $U(1)$ symmetry. This symmetry is explicitly broken by QCD instantons, which generate a nonzero axion mass and makes the axion a viable dark matter candidate~\cite{Preskill:1982cy,Abbott:1982af,Dine:1982ah}. A large experimental effort is devoted to searching for the axion, and therefore knowing the mass range of the axion is important.

The origin of the axion mass can be traced to the so-called large instanton contributions\footnote{Large instanton contributions correspond to the strong coupling regime of QCD and thus are not calculable. However, using chiral symmetry one can relate the axion mass to the equally incalculable but experimentally known pion mass, see for example \cite{diCortona:2015ldu}.} in QCD. These are contributions to the path integral that arise from instantons of size $\rho \sim 1/\Lambda_{\rm QCD}$, where $\Lambda_{\rm QCD}$ is the QCD strong coupling scale. This IR contribution dominates the integration measure over the instanton collective coordinates because the theory is asymptotically free and therefore instantons of much smaller (UV) size give negligible contributions to the non-perturbatively generated axion potential in QCD. However this implicitly assumes that the QCD coupling remains asymptotically free in the UV and QCD dynamics is not modified below the Planck scale. 

Thus in attempts to enhance the axion mass it is natural to speculate on possible UV modifications of QCD dynamics, such as those considered in Refs.~\cite{Dine:1986bg,Choi:1988sy, Rubakov:1997vp, Choi:1998ep, Berezhiani:2000gh, Hook:2014cda, Fukuda:2015ana, Gherghetta:2016fhp, Dimopoulos:2016lvn}. Of particular interest to us will be the possibility proposed in \cite{Holdom:1982ex,Holdom:1985vx,Flynn:1987rs}, where QCD is strongly coupled in the UV and thus enhances the small instanton contributions to the axion potential.  In this paper we explicitly construct such a UV modification by embedding QCD in a five-dimensional (5D) theory. In addition to the bulk QCD gluons, we identify the axion with the 5th component of a $U(1)$ gauge field, while the axion-gluon coupling arises from a 5D Chern-Simons term. 

This UV modification of QCD implies that the axion mass can now receive contributions from 5D $\it small$ instantons. The 5D instanton solution is obtained by simply uplifting the usual 4D instanton~\cite{Belavin:1975fg} to five dimensions. This gives a finite 5D action provided the extra dimension is compact (of size $\pi R$), and leads to a well-defined semiclassical expansion of the path integral around this solution~\cite{Csaki:2001zx}. The axion mass contributions can then be calculated in the perturbative limit by restricting the number of Kaluza-Klein modes. As was shown in~\cite{Poppitz:2002ac} using deconstruction, besides the usual logarithmic terms present in the effective action in the instanton density, there is a power-law term $R/\rho$ that arises from the positive frequency modes of the Kaluza-Klein gluon states. We show that this result can also be obtained by performing a fully 5D calculation of the Kaluza-Klein contributions to the effective action which is just a 5D version of 't Hooft's computation~\cite{tHooft:1976snw}. The power-law term in the effective action can be sizeable for small instantons ($\rho\ll R$), leading to a possible enhancement of the axion mass. 

This new contribution to the axion mass can be compared with the usual low-energy QCD contribution from large instantons. Interestingly, we will see that the 5D small instantons can provide the dominant contribution, but at the expense of the 5D theory being near the non-perturbative limit. In addition the enhancement is maximized only when the SM fermions are confined to the boundary of the extra dimension. Under these conditions we find that the axion mass can be enhanced by many orders of magnitude, depending on the size of the extra dimension. Since the 5D theory is near the non-perturbative limit we also consider the impact of higher dimension terms in the 5D Lagrangian, and show that they lead to noticeable but controllable effects provided the scale suppressing the higher dimension terms is smaller than the scale at which the 5D theory becomes strongly coupled. Our results for the axion mass have consequences for the experimental efforts searching for the axion (or axion-like particles, in general) with the conclusion that large regions of parameter space could remain viable.

The outline of this paper is as follows. In section~\ref{sec: pureYM} we present our 5D model for a pure Yang-Mills theory. The 4D instanton solution is then uplifted to five dimensions and shown to give a power-law term in the effective action in section~\ref{sec:5Dsmallinstantons}. The effect of including higher dimension terms 
is discussed in section~\ref{sec:highdimterms}, and the contributions to the axion mass are calculated in
section~\ref{sec:Axionmass}.  In section~\ref{sec:FermionAxionmass} we consider the fermion contributions
to the axion mass for the case of boundary fermions (section~\ref{sec:branefermions}) and bulk fermions (section~\ref{sec:bulkfermions}). In section \ref{sec:4Dmoosemodels} we compare the dynamics of 5D small instantons 
with the small instantons of 4D moose models introduced in \cite{Agrawal:2017ksf,Agrawal:2017evu}. 
Our concluding remarks are given in section~\ref{sec:conclusion}. Appendix~\ref{sec:App5D} contains the details of the 5D calculation of the Kaluza-Klein contributions to the effective action, while the calculation performed using the 4D deconstruction is summarized in Appendix~\ref{sec:4Ddeconstruction}.

\section{5D Instantons and the Axion Mass in a pure Yang-Mills theory}
\label{sec: pureYM}

We will consider a 5D spacetime $(x^\mu,y)$ where the 5th dimension, $y$ is compactified on an orbifold of size $L=\pi R$ with the QCD gauge group $SU(3)_c$ in the bulk. The bulk QCD gauge boson $A_M$ ($M=\mu,5$) will have $(+,+)$ boundary conditions for the $A_\mu$ components, while the $A_5$ components will have $(-,-)$ boundary 
conditions.\footnote{The notation $(\cdot,\cdot)$ refers to either Neumann $(+)$ or Dirichlet $(-)$ boundary conditions at $y=0$ (first entry) and $y=L$ (second entry).}
The QCD gluon is thus identified with the zero mode $A_\mu^{(0)}$. In addition, the bulk contains a $U(1)$ gauge group where the $U(1)$ gauge boson $B_M$ has $(-,-)$ boundary conditions for the $B_\mu$ components,
and $(+,+)$ boundary conditions for the $B_5$ component. This ensures that there is a massless pseudoscalar zero mode $B_5^{(0)}$ (to be identified with the axion), whereas the Kaluza-Klein (KK) scalar modes ($B_5$) are eaten by the KK $U(1)$ gauge bosons to become massive. We will start by first considering the pure YM case without any fermions.

In order to generate an  anomalous axion coupling to gluons below the compactification scale $1/R$, a bulk Chern-Simons term must also be added. The 5D Lagrangian of the $SU(3)_c\times U(1)$ theory with a Chern-Simons term is given by
\begin{equation}
S_5= -\int d^4 x \int_0^L dy\left(\frac{1}{4 g_5^2} {\rm Tr}[G_{MN}^2] +\frac{b_{CS}}{32\pi^2} \varepsilon^{MNRST} B_M {\rm Tr} [G_{NR} G_{ST}] +  \frac{1}{4 g_5^2} F_{MN}^2+\dots \right)~,
\label{eq:5Daction}
\end{equation}
where $G_{MN} (F_{MN}) $ is the gluon ($U(1)$) field-strength tensor, $b_{CS}$ is a dimensionless constant and we have equally normalized the non-Abelian and Abelian gauge fields with $g_5$ the (dimensionful) gauge coupling. The 5D gauge theory has a UV cutoff $\Lambda_5$ whose maximum value occurs where the theory becomes strongly coupled, $g_5^2 \Lambda_5/(24 \pi^3) \sim1$. Higher dimension terms in the Lagrangian are expected to be suppressed by $\Lambda_5$, and for now they have been neglected in (\ref{eq:5Daction}). Later we will see that they can have an important effect on the 5D instanton. Note that a $\theta$ term is not allowed in the 5D action (\ref{eq:5Daction}) due to Lorentz invariance, but can be present on the 4D boundaries. However the $U(1)$ symmetry in 
(\ref{eq:5Daction}) can be used to eliminate these boundary $\theta$ terms. This is one of the distinguishing features of our 5D model compared to that of the 4D moose models~\cite{Agrawal:2017ksf, Agrawal:2017evu} where there is  a theta angle for each  $SU(3)$ gauge group, and therefore one has to also introduce an axion at each site.

Upon compactification we obtain the effective 4D Lagrangian
\begin{equation}
 S_4= \int d^4 x~\left(\frac{1}{4 g_s^2} {\rm Tr}[G_{\mu\nu}^{2}] + \frac{1}{32\pi^2}
 \frac{a}{f} {\rm Tr}[G_{\mu\nu} 
{\widetilde G}^{\mu\nu}] +  \frac{1}{2} (\partial_\mu a)^2+\dots \right)~,
\label{eq:4DLag}
\end{equation}
where $g_s$ is the 4D QCD gauge coupling, $G_{\mu\nu} \equiv G_{\mu\nu}^{(0)}$ is the QCD gluon field strength tensor, $a\equiv B_5^{(0)}/g_s$ is the axion\footnote{Alternatively the axion could be a localized boundary field that couples to ${\rm Tr}[G_{\mu\nu}  {\widetilde G}^{\mu\nu}]$. Our analysis also applies in this case.}, and the couplings are identified as
\begin{equation}
     \frac{1}{g_s^2}\equiv \frac{L}{g_5^2},\qquad \frac{1}{f}\equiv  
       b_{CS} g_s L~.
     \label{eq:couplings}
\end{equation}

We will assume that the 5D cutoff of the model, $\Lambda_5$ lies at or below 
the  strong coupling scale $\sim 24 \pi^3/g_5^2$. Using (\ref{eq:couplings}) this translates into the limit
\begin{equation}
         \Lambda_5 R \lesssim \frac{6\pi}{\alpha_s}~,
         \label{eq:Lam5Rlimit}
\end{equation}
where $\alpha_s=g^2_s/(4\pi)$ and for an orbifold, $L=\pi R$. 
Thus for $\alpha_s \sim 0.1$ we obtain $\Lambda_5 R \lesssim 200$.

\subsection{5D Small Instantons}
\label{sec:5Dsmallinstantons}

The extra dimension provides a UV modification of QCD at the scale $1/R\gg \Lambda_{\rm QCD}$, where 
$\Lambda_{\rm QCD}\simeq 300$~MeV is the QCD strong scale. It is thus possible that instantons of size  
$\lesssim R$ can give large contributions to the axion mass. Let us consider first how the instanton calculus is modified in the 5D pure YM $SU(3)$ theory. The compactified 5D theory admits the following instanton solution in Euclidean space:
\begin{equation}
      A_\mu^a(x,y ) =A_\mu^{(I)a}(x), \qquad A_5^a(x,y ) = 0,
      \label{eq:5Dinstantonsoln}
\end{equation}
where 
\begin{equation}
  A_\mu^{(I) a}(x) = \frac{2\, \eta^a_{\mu\nu}(x-x_0)_\nu}{(x-x_0)^2+\rho^2}~,
          \label{eq:4Dinstanton}
\end{equation}
is the 4D instanton configuration~\cite{Belavin:1975fg} in the regular gauge with center $x_0$ and size $\rho$.
The tensors $\eta^a_{\mu\nu}$ are the group-theoretic 't Hooft eta-symbols~\cite{tHooft:1976snw} and $a$ denotes the gauge isospin index. The 5D instanton solution (\ref{eq:5Dinstantonsoln}) can be simply thought of as wrapping the 4D solution (\ref{eq:4Dinstanton}) around the compact dimension.

Note that in the deconstructed version of \cite{Poppitz:2002ac} the above 5D solution corresponds to a multi-instanton configuration with winding numbers $(1,1,\dots,1)$.  Importantly, it does not appear to be the continuum limit of the 4D instantons used in the moose model of \cite{Agrawal:2017ksf} that corresponds to the combination of $(1,0,0\dots,0) + (0,1,0\dots,0) + (0,0,1\dots,0) +\dots$. This latter configuration would correspond to instantons localized in the bulk i.e. $A_\mu(x,y ) =A_\mu^{(I)}(x)\delta (y)$. However this is not a solution of the 5D equations of motion, and therefore the  4D moose model of \cite{Agrawal:2017ksf} does not reconstruct to a 5D theory.

\begin{figure}[t]
\centering
\includegraphics[width=0.7\textwidth]{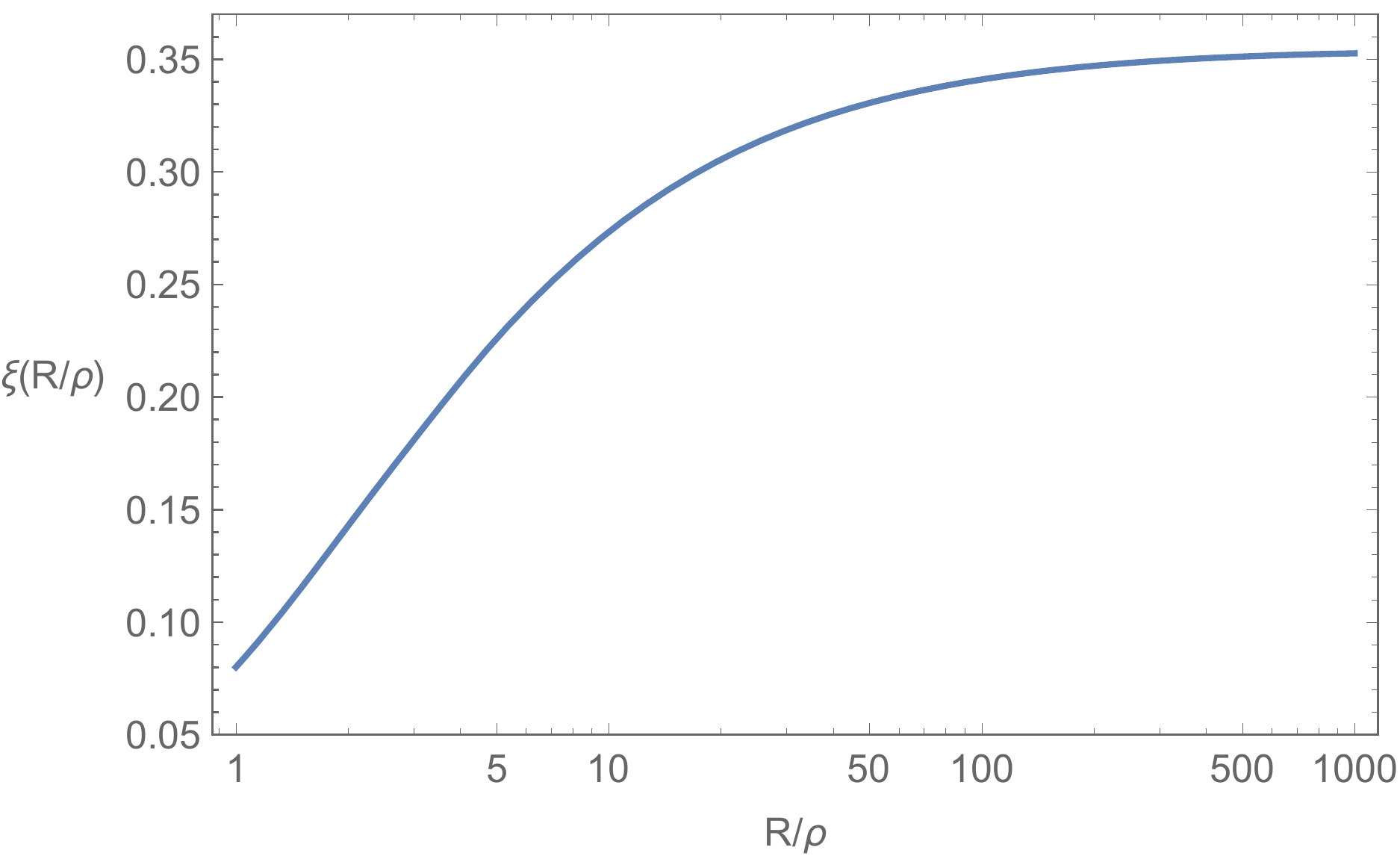}
\caption{\it $\xi$ as a function of $R/\rho$.}
\label{fig:xi}
\end{figure}

The 5D instanton solution (\ref{eq:5Dinstantonsoln}) minimizes the action $(\ref{eq:5Daction})$  to give (ignoring for now the axion terms)
\begin{equation}
         S_5^{(I)}= \frac{8\pi^3 R}{g_5^2} = \frac{2\pi}{\alpha_s}~,
          \label{eq:S5extremum}
\end{equation}
where we have used the relation (\ref{eq:couplings}) with $L=\pi R$. To obtain the contribution of the instanton
to the partition function  we must  also consider the fluctuations of the 5D gauge fields about the instanton solution (\ref{eq:5Dinstantonsoln}). This means not only including the gluon fluctuations but also the KK contributions.
The details of this  calculation are given in Appendix~\ref{sec:App5D}, and the final result for a pure Yang-Mills $SU(N)$ theory is presented in (\ref{inst5dSUN}). For $SU(3)$ the result is 
\begin{equation}
      \int_{1/\Lambda_5}^R \frac{d\rho}{\rho^5} \,C[3]\left(\frac{2\pi}{\alpha_s(1/R)}\right)^6 e^{-S_{\rm eff}}
      \equiv \frac{K}{R^4}~,
       \label{eq:inst5dSU3}
\end{equation}
where $C[3]\simeq 1.5\times 10^{-3}$, and the effective action is given by
\begin{equation}
S_{\rm eff}={\frac{2\pi}{\alpha_s(1/R)}-3\xi(R/\rho) \frac{R}{\rho}+b_0\ln\frac{R}{\rho}}~,
\label{eq:SeffYM}
\end{equation}
where $\alpha_s(1/R)$ is the YM coupling evaluated at $1/R$ (see \eqref{gaugeR} for the exact definition)
and  $b_0 = 11$ (the pure QCD $\beta$ function coefficient) is the contribution from the gauge boson zero modes. The function $\xi(R/\rho)$ is plotted in Figure~\ref{fig:xi}. The quantity $K$ that appears on the r.h.s. of \eqref{eq:inst5dSU3} is a dimensionless factor resulting from evaluating the integral in \eqref{eq:inst5dSU3}. Note that since we are only considering the effect of 5D small instantons, the integration region in (\ref{eq:inst5dSU3}) is limited to $1/\Lambda_5 \leqslant \rho \leqslant  R$. The dependence  of  the lower limit of integration on $1/\Lambda_5$ can make the contribution very sensitive to the UV completion details. This will be further discussed in Section~\ref{sec:highdimterms}.

The result (\ref{eq:inst5dSU3}) reveals a new, interesting feature. There is a power-law term $(R/\rho)$ in the exponent arising from the KK modes with a positive coefficient, $\xi(R/\rho)>0$, which now causes the integral over the instanton size $\rho$ to receive a large  contribution from the  $\it small$ instantons of size, $\rho\sim 1/\Lambda_5$.  
As we will show, in some parameter regions this contribution can overcome the IR contribution  
dominated by large instantons of order $\rho\sim 1/\Lambda_{\rm QCD}$.

An approximate expression for the dimensionless factor $K$ on the r.h.s. of (\ref{eq:inst5dSU3}) can be obtained by 
evaluating the integral in \eqref{eq:inst5dSU3} and using the fact that $\xi(R/\rho) \sim 1/3$ for $R/\rho \gtrsim 20$. This gives
\begin{equation}
K\simeq C[3] \left(\frac{2\pi}{\alpha_s(1/R)}\right)^6 (\Lambda_5 R)^{3-b_0} e^{-\frac{2\pi}{\alpha_s(1/R)}+\Lambda_5 R}
= C[3] \left(\frac{2\pi}{\alpha_s(1/R)}\right)^6 
\frac{e^{-\frac{2\pi}{\alpha_s(1/R)}+\Lambda_5 R}}{(\Lambda_5 R)^8}\,.
  \label{eq:Kconstant}
\end{equation}
Thus we see that for sufficiently large $\Lambda_5 R$, the power-law term in the effective action \eqref{eq:SeffYM}
 leads to an exponential enhancement that can overcome the suppression from $e^{-2\pi/\alpha_s(1/R)}$ to give a  UV-dominated contribution to the integral in \eqref{eq:inst5dSU3}. Note that for the calculation to be reliable, $\Lambda_5 R$ cannot saturate the bound (\ref{eq:Lam5Rlimit}), otherwise higher-loop corrections in the instanton background will be equally important. Furthermore the fact that the contribution (\ref{eq:Kconstant}) is cutoff dependent suggests that higher dimension terms in the 5D Lagrangian are also important and should be considered.

\subsection{Higher dimension  terms}
\label{sec:highdimterms}

To study the impact of higher dimension terms, we consider the addition of the following dimension six term to the 5D action:
\begin{equation}
   \Delta   S_5 = -\frac{1}{4 g_5^2} \int d^4 x \int_0^{L} dy \,
     \frac{c_6}{\Lambda_5^2}\, {\rm Tr}\, G_{MN}\Box G^{MN}~,
      \label{eq:5Dhighdim}
\end{equation}
where $c_6$ is a dimensionless constant. We will assume that $c_6>0$ so that it stabilizes the instanton action. Substituting (\ref{eq:5Dinstantonsoln}) into (\ref{eq:5Dhighdim}) and performing the 5D integration leads to
\begin{equation}
      S_{\rm eff} = \frac{2\pi}{\alpha_s}  +\frac{3\pi}{\alpha_s} \frac{c_6}{(\Lambda_5\rho)^2} -3\xi(R/\rho) \frac{R}{\rho}+\dots~,
      \label{eq:sub5Daction}
\end{equation}
where the logarithmic term in (\ref{eq:SeffYM}) has been neglected.
Note that the instanton solution (\ref{eq:5Dinstantonsoln}) is itself modified by the order $1/\Lambda_5^2$ terms in (\ref{eq:5Dhighdim}). However these corrections lead to subleading terms of order $1/(\Lambda_5\rho)^4$ in $S_{\rm eff}$, and can be neglected. 
Whereas $S_{\rm eff}$ is extremized near the UV size $1/\Lambda_5$ when $c_6=0$, the inclusion of the higher dimension term in (\ref{eq:sub5Daction}) instead leads to an extremum
\begin{equation}
      \frac{1}{\rho_*} \simeq \frac{3}{c_6} \xi(R/\rho) \left(\frac{g_5^2 \Lambda_5}{24\pi^3}\right) \Lambda_5~,
\label{eq:ext}
\end{equation}
where the $\rho$ dependence in $\xi(R/\rho)$ has been neglected since it is approximately constant for $\rho\ll R$.
As long as the theory is perturbative at the cutoff, $g_5^2 \Lambda_5/(24\pi^3)\ll 1$, the extremum condition (\ref{eq:ext}) implies $\rho_* \gg 1/\Lambda_5$, and therefore the contribution  (\ref{eq:inst5dSU3}) is dominated by instantons of size $\rho_*$. As alluded to earlier, this means that the instanton size is effectively cutoff at $\rho_*$, and the factor $K$ is approximately given by the expression (\ref{eq:Kconstant}) with $\Lambda_5$ replaced by $1/\rho_*$. Of course there is no need to rely on the approximate expression, and one can simply perform the numerical integration in (\ref{eq:inst5dSU3}) to obtain the exact factor $K$. To reiterate the salient point, the integral in (\ref{eq:inst5dSU3}) with the higher dimension term (\ref{eq:5Dhighdim}) included, is dominated by instantons of size $\rho_*$ where the 5D theory remains perturbative, and therefore contributions from instantons of size $1/\Lambda_5$ are suppressed. Furthermore, higher dimension  terms (beyond those of (\ref{eq:5Dhighdim}))  can be neglected, as they are suppressed by higher powers of $\frac{1}{\Lambda_5 \rho_*}\ll 1$,  and the calculation remains under theoretical control.

\subsection{5D Small Instanton Corrections to the Axion Mass}
\label{sec:Axionmass}

To calculate the contribution to the axion mass from the enhanced instanton density, we next
add the axion field. For a constant background axion field, $a$, in the instanton background 
(\ref{eq:5Dinstantonsoln}) we obtain
\begin{equation}
       i \frac{a}{f}\frac{1}{32\pi^2}\int d^4 x\,{\rm Tr}[G^{(I)}_{\mu\nu}  {\widetilde G}^{(I)\mu\nu}]=i\frac{a}{f}\,,
\end{equation}
where the winding number is one. The effective action (\ref{eq:SeffYM}) is then modified
by replacing $S_{\rm eff} \rightarrow S_{\rm eff} -ia/f$. Summing over 
both instanton and anti-instanton contributions in the dilute instanton gas approximation~\cite{Callan:1977gz,Vainshtein:1981wh},  leads to 
\begin{eqnarray}
      Z&=& \sum_{n,{\bar n}=0}^\infty \frac{1}{n!{\bar n}!} \prod_{k=1}^{n}\left( \int d^4x_k \frac{K}{R^4} e^{-i\frac{a}{f}}\right) \prod_{{\bar k}=1}^{\bar n}\left( \int d^4x_{\bar k} \frac{K}{R^4} e^{i\frac{a}{f}}\right)\nonumber \\
      &=& \exp\left[ 2 \frac{K}{R^4} \int d^4 x \cos\left(\frac{a}{f}\right)\right]~.
\end{eqnarray}
We conclude that the contribution of the 5D small instantons to the axion mass is
\begin{equation}
       m_a^2 = \frac{2K}{f^2R^4}~,
\end{equation}
where $K$ is defined in (\ref{eq:inst5dSU3}) and $f$ is given in (\ref{eq:couplings}).  

\begin{figure}[t]
\centering
\includegraphics[width=0.8\textwidth]{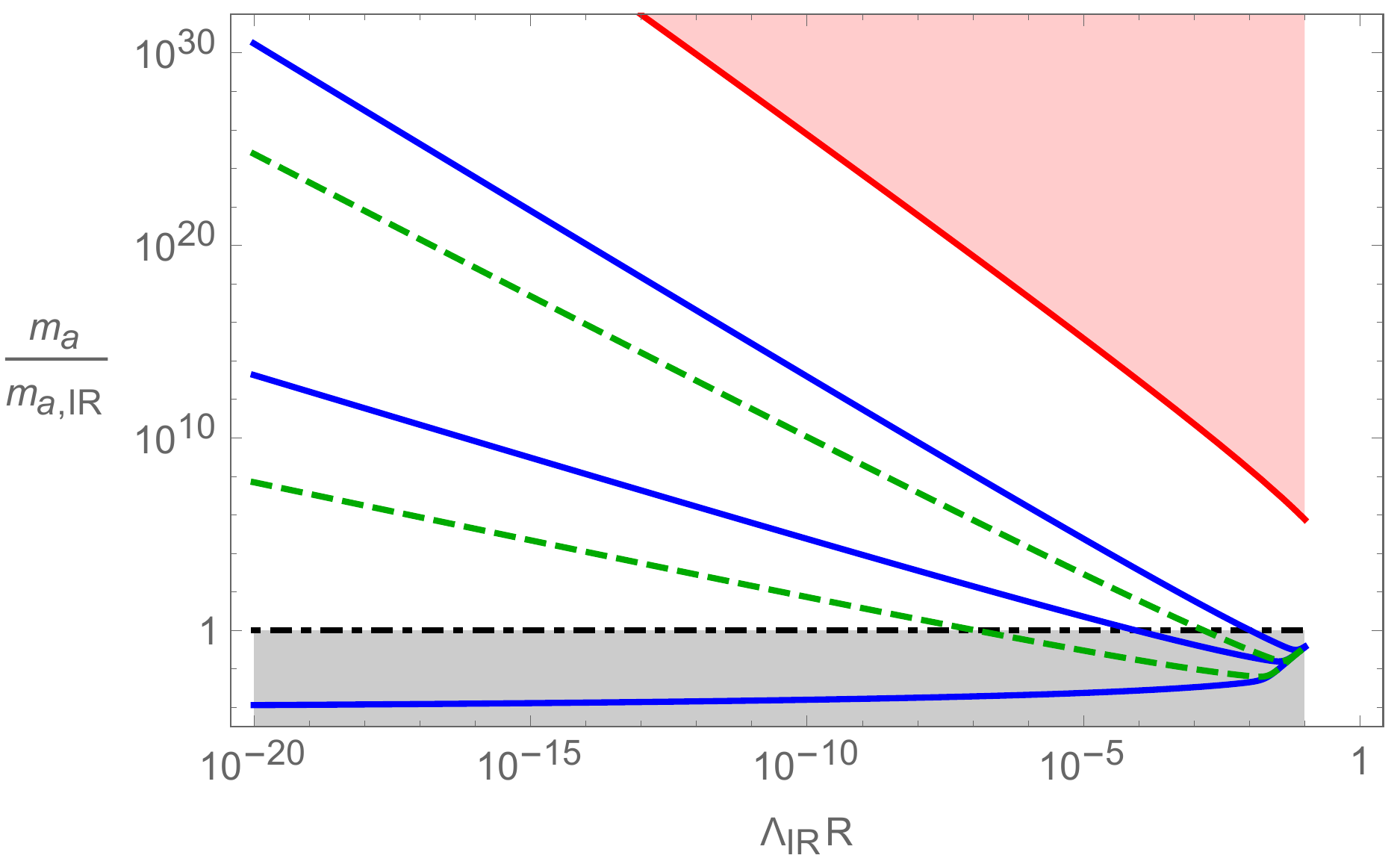}
\caption{\it The axion mass ratio for the pure YM case as a function of $\Lambda_{IR} R$ for various contours of $\epsilon =(0.3,0.25,0.2)$ (top to bottom). The solid blue lines are the exact numerical integration results using (\ref{eq:inst5dSU3}) with no higher dimension terms ($c_6=0$).  The green dashed line represents the addition of the higher dimension term (\ref{eq:5Dhighdim}) with $c_6=0.5$ and $\epsilon = 0.52 (0.47)$ for the upper (lower) line. The red line represents the maximum enhancement using $m_{a,5}$ that would occur at the 5D strong-coupling limit.}
\label{fig:YMaxionratio}
\end{figure}

To understand the importance of the 5D small instanton contributions to the axion mass, it is  
instructive to compare these UV contributions with the IR (large) instanton contributions, which for the pure YM case are estimated to be $m_{a,IR}^2 \sim \Lambda_{IR}^4/f^2$, and
\begin{equation}
\Lambda_{IR} = \frac{1}{R} e^{-\frac{2\pi}{b_0 \alpha_s(1/R)}}~,
\label{eq:LamIR}
\end{equation}
is defined as the IR scale at which the gauge coupling becomes strong.
The axion mass ratio is then 
\begin{eqnarray}
    \frac{m_a}{m_{a,IR}} &\simeq&
    \sqrt{2C[3]}\left(\frac{2\pi}{\alpha_s(1/R)}\right)^3  \frac{1}{(\Lambda_5 R)^4}  e^{-\frac{1}{2} \left( \frac{7}{11} \frac{2\pi}{\alpha_s(1/R)} -\Lambda_5 R\right)}\,,
     \label{eq:approxmassratio} \label{eq:axionmassratioA}\\
    &=& \sqrt{2C[3]}\left(\frac{2\pi}{\alpha_s(1/R)}\right)^3  \frac{(\Lambda_{IR} R)^{7/2}}{(\Lambda_5 R)^4}  e^{\frac{1}{2} \Lambda_5 R}\,,
   \label{eq:axionmassratio}
\end{eqnarray}
where we have used the approximation (\ref{eq:Kconstant}). It is clear that $m_a\gtrsim m_{a,IR}$ when the exponent in (\ref{eq:approxmassratio}) is $\gtrsim 0$. By writing 
\begin{equation}
\Lambda_5 R =\frac{6\pi \epsilon }{\alpha_s(1/R)}\,,
\label{eq:Lam5Reps}
\end{equation}
where $\epsilon\lesssim 1$ using the perturbativity condition (\ref{eq:Lam5Rlimit}),
we obtain a positive exponent  in \eqref{eq:axionmassratioA}
for $\epsilon \gtrsim 7/33\simeq 0.21$.
This shows that a large contribution to the axion mass from 5D small instantons requires 
values of $\Lambda_5 R$ near the non-perturbative limit. In Figure~\ref{fig:YMaxionratio} 
we plot the  mass ratio ${m_a}/{m_{a,IR}}$ using the exact result  for $K$,  and taking
 different values of the perturbativity parameter $\epsilon$ defined in \eqref{eq:Lam5Reps}. We see 
that in order to have a large contribution to the axion mass we must be quite close 
to the non-perturbative limit $\epsilon\sim 1$.

In the limit in which the theory is strongly coupled at $\Lambda_5$ ($\epsilon\sim 1$)
and the instanton contributions are  dominated by instantons of size $1/\Lambda_5$, 
we cannot reliably calculate the 5D instanton contribution to the axion mass.
However a naive dimensional analysis estimate gives 
\begin{equation}
      m_{a,5}^2 \sim \frac{\Lambda_5^4}{f^2}\,, 
\end{equation}
up to an order-one constant. This estimate corresponds to the maximum value of the axion mass from 5D instantons, and is shown as a red  line in Figure~\ref{fig:YMaxionratio}.

\begin{figure}[t]
\centering
\includegraphics[width=0.35\textwidth]{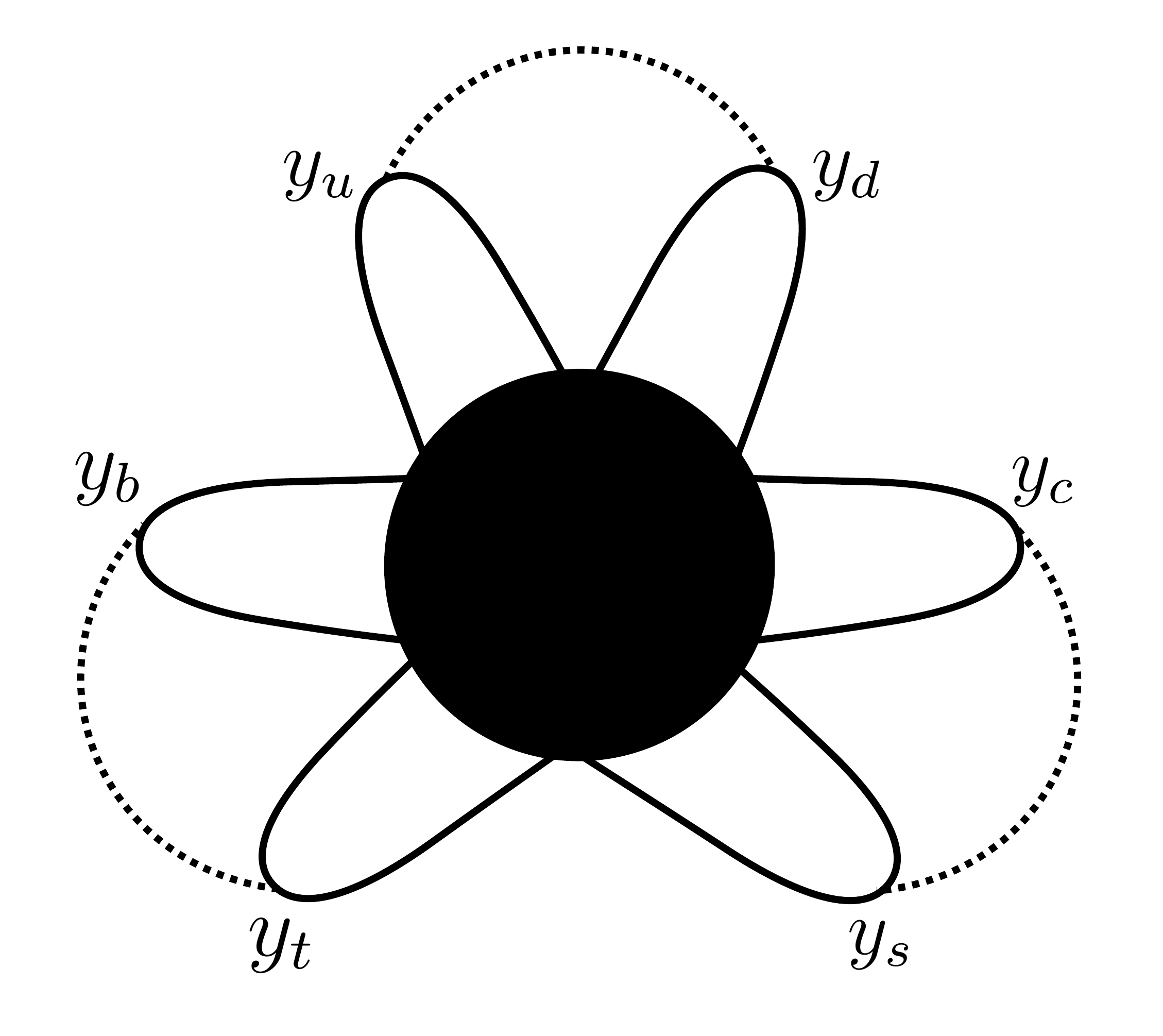}
\caption{\it Instanton vacuum diagram arising from closing the SM fermion legs with  Higgs loops.}
\label{fig:instantondiagram}
\end{figure}

\section{Fermion contributions and QCD axion mass}
\label{sec:FermionAxionmass}

So far we have considered a pure YM theory. We next introduce fermions in order to discuss QCD in the SM. In five dimensions the fermions are not chiral, and thus to reproduce the SM matter content one must impose boundary conditions that project out the unwanted Weyl components of the Dirac fermions. Alternatively, one can assume that all SM matter fields are confined to the boundary of the extra dimension. Indeed, the assumption that all SM fermions are on the boundary is essential for our purposes: as explained in \cite{Poppitz:2002ac} the sign of the $R/\rho$ term responsible for the enhancement of the small instanton contribution flips if the number of bulk fermions is sufficiently large. As shown in Appendix~\ref{sec:App5D}, in the 5D orbifolded theory this happens when $N_f>9N/4$. Thus in section \ref{sec:branefermions} we will first consider boundary fermions (identified with SM matter fields), and calculate the small instanton contributions to the axion mass assuming the 5D theory is in the perturbative regime. We will also estimate the axion mass when the UV cutoff is at the strong-coupling scale $\sim 24\pi^3/g_5^2$.
Bulk fermions will then be discussed in section \ref{sec:bulkfermions}.

\subsection{Boundary fermions}
\label{sec:branefermions}

If the SM fermions (as well as the Higgs field) are on the boundary and $SU(3)$ is in the bulk, 
the instanton integral remains the same as (\ref{eq:inst5dSU3}) except that $b_0=7$ (to account for the fermion zero modes). Furthermore, fermion zero modes would naively lead to an extra suppression factor in the instanton density of the form $(\rho m_f)^{N_f}$ where $m_f$ are the fermion masses and $N_f$ the number of flavors. However since the fermion masses in the SM arise from a Higgs mechanism, the fermion legs in an instanton vacuum diagram can be closed with a Higgs loop (see Figure~\ref{fig:instantondiagram}). Thus the suppression is only proportional 
to the Yukawa couplings and loop factors, namely:
\begin{equation}
      \kappa_f = \frac{y_u}{4\pi} \frac{y_d}{4\pi} \frac{y_c}{4\pi} \frac{y_s}{4\pi} \frac{y_t}{4\pi} \frac{y_b}{4\pi} \approx 10^{-23}~,
      \label{eq:kappaf}
\end{equation}
where $y_{u,d,c,s,t,b}$ are the SM Yukawa couplings. This is one of  the ingredients leading to the enhancement of the axion mass in the 4D moose models of~\cite{Agrawal:2017ksf,Agrawal:2017evu}, as well as in our 5D model. 

\begin{figure}[t]
\begin{center}
\includegraphics[width=.8\textwidth]{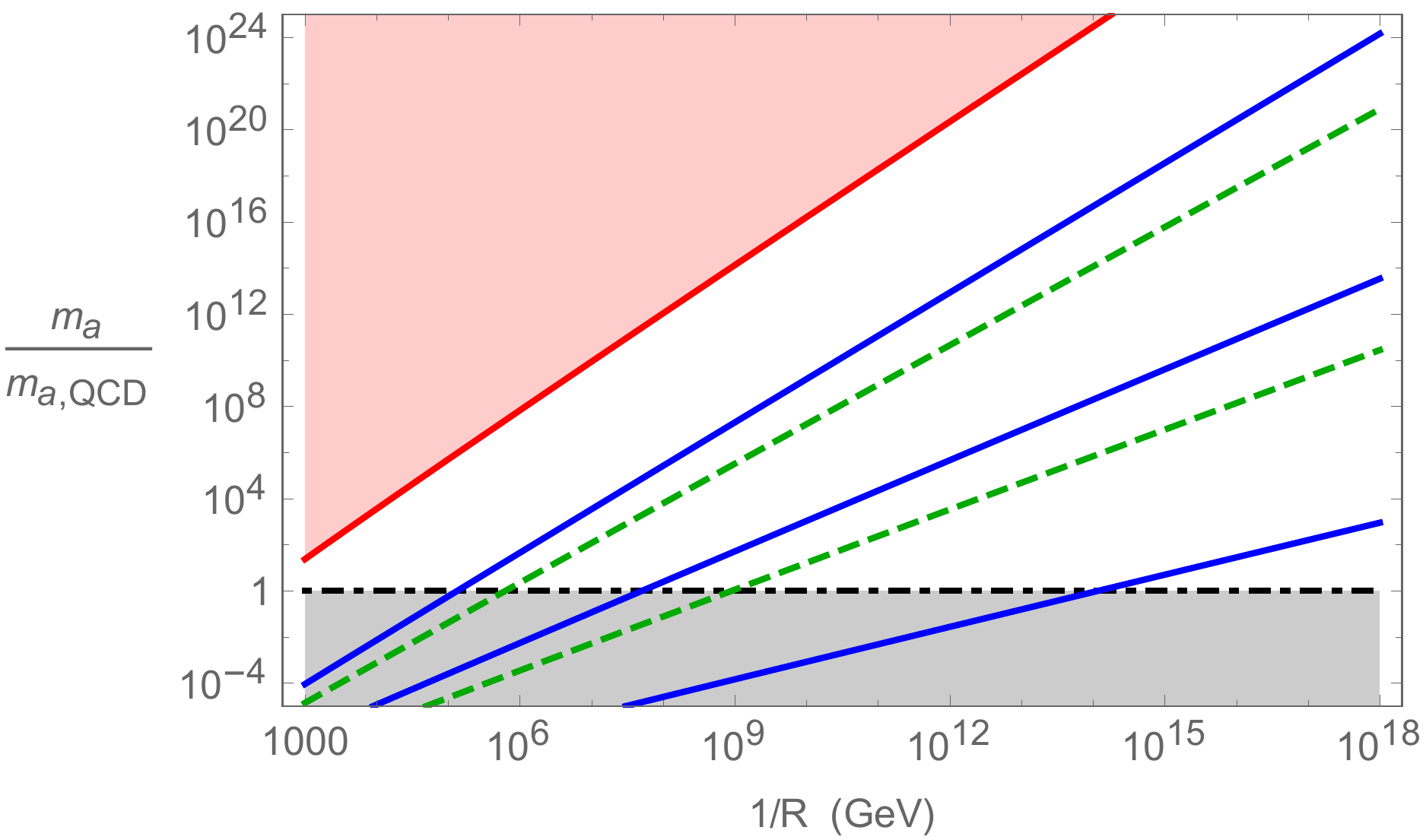}
\caption{\it The axion mass ratio for the boundary fermions case (assuming $\alpha_s(m_Z)=0.118$), as a function of $1/R$ for various contours of $\epsilon =(0.3,0.25,0.2)$ (top to bottom). The solid lines are the exact results obtained from a numerical integration of (\ref{eq:inst5dSU3}) and no higher dimension terms ($c_6=0$).  The green dashed line represents the addition of the higher dimension term (\ref{eq:5Dhighdim}) with $c_6=0.5$ and $\epsilon = 0.52\, (0.47)$ for the upper (lower) line. The red line depicts the maximum enhancement in the strong coupling limit using $m_{a,5f}$.}
\label{fig:axionmassratio}
\end{center}
\end{figure}

With the introduction of fermions, the axion mass low-energy contribution can be unambiguously determined from QCD chiral perturbation theory to be \cite{Weinberg:1977ma,DiVecchia:1980yfw},
\begin{equation}
          m_{a,QCD}^2 = \frac{m_u m_d}{(m_u+m_d)^2} \frac{m_\pi^2 f_\pi^2}{f^2}\,,
          \label{eq:axionfermionmass}
\end{equation}
where $m_\pi \simeq 135$~MeV, $f_\pi \simeq 92$ MeV, and $m_u/m_d \simeq 0.46$.
Using the result (\ref{eq:Kconstant}) with $b_0=7$ and including the factor (\ref{eq:kappaf}), the axion mass ratio becomes:
\begin{equation}
     \frac{m_a}{m_{a,QCD}}  \simeq \sqrt{2\kappa_f C[3]} \left(\frac{2\pi}{\alpha_s(1/R)}\right)^3   \frac{(m_u+m_d)}{\sqrt{m_u m_d}}\frac{1}{ m_\pi f_\pi R^2} 
     \frac{e^{-\frac{1}{2}\left(\frac{2\pi}{\alpha_s(1/R)}-\Lambda_5 R\right)}}{(\Lambda_5 R)^{\frac{1}{2}(b_0-3)}}\,.
     \label{eq:boundaryfermionratio}
\end{equation}
Here we have not considered higher dimension terms, so the path integral is dominated by instantons of 
size  $\rho_*\sim 1/\Lambda_5$. As discussed in sec.~\ref{sec:highdimterms},  the presence of higher dimension terms can increase $\rho_*$, making the result less dependent on the cutoff. Note that in (\ref{eq:boundaryfermionratio}) the chiral suppression factor $\kappa_f$ is mitigated by the fact that $\alpha_s$ runs slower towards the UV due to the SM fermions, and therefore $\alpha_s(1/R)$ is larger, implying  that the exponential suppression is smaller. Using (\ref{eq:Lam5Reps}) and approximating the pion and quark masses with the QCD scale, we obtain a positive  exponent in \eqref{eq:boundaryfermionratio} for $\epsilon \gtrsim 0.14$. The exact numerical result for ${m_a}/{m_{a,QCD}}$ is plotted in Figure~\ref{fig:axionmassratio}, where a sizeable enhancement can be seen that depends sensitively on $\epsilon$. An enhancement at low compactification scales requires larger values of $\epsilon$.

The maximum possible enhancement occurs when the 5D theory is strongly coupled at $\Lambda_5$. In this case the naive dimensional analysis estimate for the axion mass with fermion contributions then becomes
\begin{equation}
       m_{a,5f}^2 \sim \kappa_f  \frac{\Lambda_5^4}{f^2}~,
       \label{eq:fermLam5}
\end{equation}
where the suppression  factor $\kappa_f$ defined in \eqref{eq:kappaf} has been included, since we assume that there are no other sources of chiral breaking in the 5D model beyond the SM Higgs Yukawas. 

The fermion suppression $\kappa_f$ in (\ref{eq:fermLam5}) could actually be removed if we relax this assumption and consider extra heavy Higgs fields coupled to the SM quarks  with order one Yukawa couplings~\cite{Dine:1986bg}. However this comes at the expense of possibly introducing new CP phases in the heavy Higgs couplings that could spoil the axion solution to the strong CP problem. Even if these heavy Higgs fields are introduced (and for some reason do not introduce new phases) the suppression cannot be entirely removed because there is a maximum value for the 5D instanton contribution to the axion mass. This arises from the fact that 5D small instantons can also contribute to the up-quark Yukawa coupling (for instance, $y_u^{(I)}$) that cannot be larger than the experimental value $y_u\sim m_u/v$ (unless we tune $y_u^{(I)}$ with the SM Yukawa coupling). By closing the  up-quark and Higgs legs of this instanton  contribution, we  can then obtain a two-loop contribution to the  axion mass of order  $m_{a}^2 \sim \frac{y_u^{(I)} y_u}{(16\pi^2)^2} \frac{\Lambda_5^4}{f^2} \lesssim \frac{y_u^2}{(16\pi^2)^2} \frac{\Lambda_5^4}{f^2}$. Absent a fine tuning between the tree-level ($y_u$) and the instanton ($y_u^{(I)}$) contributions this would then represent the maximum possible enhancement allowed for alleviating the $\kappa_f$ suppression in (\ref{eq:fermLam5}).

Finally one may ask why the result of our 5D instanton calculation is interpreted as an additional contribution 
relative to the low-energy expression \eqref{eq:axionfermionmass}, rather than being merged into it.
The point is that there are different symmetry breaking parameters that control \eqref{eq:kappaf} and \eqref{eq:axionfermionmass}. The chiral perturbation theory result \eqref{eq:axionfermionmass} vanishes in the limit of vanishing quark masses, $m_{u,d}\to 0$, while our instanton contribution  to the axion mass squared is proportional to the Yukawa couplings $y_{u,d}$ (as seen in \eqref{eq:kappaf}). In fact the instanton result does not vanish in the limit of zero Higgs vacuum expectation value as can be inferred from the diagram in Figure~\ref{fig:instantondiagram}.
The Yukawa couplings explicitly break the axial $U(1)_A$ symmetry, and this effect is combined with the anomalous $U(1)_A$ breaking by the instanton to give the total instanton contribution depicted in Figure~\ref{fig:instantondiagram}. On the other hand the chiral perturbation theory contribution to the axion mass results from current quark masses which formally is a different source of explicit $U(1)_A$ and the $SU(2)$ chiral symmetry breaking. In summary there are two independent contributions to the axion mass: one proportional to $m_u m_d$ 
and the other proportional to $y_u y_d$, and they should both be taken into account.

\subsection{Bulk fermions}
\label{sec:bulkfermions}

Next we consider the fermions propagating in the 5th dimension. For each chiral SM field we need to introduce a Dirac fermion in the 5D bulk. Thus assuming $N_f$ quark flavors there are $2N_f$ Dirac fermions.  The QCD contribution from the gauge boson and fermion zero modes still gives $b_0=7$, but now there is also a contribution from the massive Dirac KK fermions. Using the results in Appendix~\ref{sec:App5D} we see that for the $SU(3)$ case, we have \eqref{eq:SeffYM} with the replacement
\begin{equation}
3\xi(R/\rho) \to  (3-4N_f/3)\xi(R/\rho)~.
\end{equation}
Compared to the pure YM case given in (\ref{eq:inst5dSU3}), the enhancement in the exponential factor $e^{-S_{\rm eff}}$  from the power-law $R/\rho$ term is now reduced as the number of flavors in the bulk increases.
In fact for $N_f=3$ the sign of the power-law $R/\rho$ term flips, which now suppresses the 5D instanton contribution. Furthermore, the fermion zero modes again lead to a suppression in the instanton vacuum diagrams 
due to Yukawa couplings and Higgs loops (assuming the Higgs is confined to the boundary). Therefore generically the best possible case for an axion mass enhancement occurs when there are no bulk fermions. 

However we would like to point out that the  introduction of bulk fermions  can  increase  the size of the dominant instantons, $\rho_*$, and therefore make the 5D  calculation less dependent on the cutoff. Indeed, if bulk fermions have large boundary  localized mass terms, $m_B\gg 1/R$, then their KK modes will not contribute to the effective instanton action at scales between $1/R$ and $m_B$. On the other hand, these heavy fermion KK modes will contribute to the effective instanton action at scales above $m_B$. As a result the effective action will have an extremum at  $1/\rho_*\sim m_B$ that can be chosen to be smaller  than the strong scale $24\pi^3/g_5^2$.

\subsection{Relation to 4D moose models}
\label{sec:4Dmoosemodels}

The contribution of small instantons from a compactified 5th dimension shares some features with the 4D moose model~\cite{Agrawal:2017ksf,Agrawal:2017evu}. In moose models the enhanced contributions of small instantons arise due to a high index of embedding of a QCD instanton into the gauge group of the microscopic theory \cite{Csaki:2019vte}. Indeed, in the simplest moose model the QCD gauge group, $SU(3)_c$, arises as a diagonal subgroup of a larger product group, $SU(3)_1\times SU(3)_2$ at some UV scale, with a theta angle and axion at each site. The QCD instanton of the two-site moose model corresponds to a multi-instanton configuration of the microscopic theory (specifically, the $(1,1)$ configuration). In other words, the small instantons in the broken gauge group factors correspond to ``fractional'' instantons of QCD, and their weights, $\Lambda_1^{b_1}$ and $\Lambda_2^{b_2}$, are large compared to the weight of the QCD instanton contribution, $\Lambda_{\rm QCD}^b$ \cite{Csaki:2019vte} (where $b_1$, $b_2$, and $b$ are the $\beta$ function coefficients of the corresponding gauge groups). As a result, the small instanton contributions to the axion masses in the broken factors are enhanced. Since the lightest mass eigenstate of these two axions plays the role of the QCD axion, this may be heavier than predicted by QCD alone. The detailed calculation in Ref.~\cite{Csaki:2019vte}  shows that the small instanton enhancement is not sufficiently large in a two-site model\footnote{This is a consequence of interactions between the Higgs field responsible for breaking the product gauge group, and the small instantons in the broken gauge group factors \cite{Csaki:2019vte, Fuentes-Martin:2019bue}.} but could be significant in $k$-site models with $k\ge 3$. Increasing the number of sites invokes the analogy between moose models and the deconstructed description of 5D theories. 

It is important to note that there is a significant qualitative difference between multi-axion moose models and truly 5D theories. In fact, the moose models of \cite{Agrawal:2017ksf,Agrawal:2017evu} do not have a 5D continuum limit because they have more than one axion and theta angle. Furthermore in truly 5D theories, the enhancement of small instanton contributions is not a consequence of a non-trivial index of embedding. Indeed, in the fully 5D theory, the gauge group is $SU(3)$ both in the UV and IR, and therefore it is obvious that the index of embedding for both small and large instantons is the same. Instead, the instanton action has two minima -- one in the IR where QCD becomes strong and another in the UV where the 5D theory becomes strong. It is the existence of this second minimum in the instanton action that is responsible for the axion mass enhancement.

\section{Conclusion}
\label{sec:conclusion}

We have shown that if QCD gluons propagate in a 5th dimension at high energies, then the effective action receives
a  power-law term  $R/\rho$ due to the positive frequency modes of the Kaluza-Klein states arising from the uplifted 4D instanton solution. This power-law term can cause the 5D small instanton contributions to the axion mass to dominate over the large instanton contribution in the IR, and therefore enhance the axion mass. However as shown in Fig.~\ref{fig:YMaxionratio}, the effects are sizeable only near the non-perturbative limit of the 5D theory.
Therefore higher loop contributions could change our conclusions. We have also considered the inclusion of higher dimension terms and shown that they  lead to an extremum in the effective action at instanton sizes $\rho_* \gg 1/\Lambda_5$. In this case the dominant contribution to the axion mass  arises from  small instantons  of sizes larger than the cutoff.

When the Standard Model fermions are included there is a suppression in the axion mass proportional to the product of Yukawa couplings. The axion mass can still receive a sizeable enhancement from the 5D small instantons 
provided the fermions are confined to the boundary. Otherwise if fermions propagate in the bulk, the KK fermion
contribution reduces the coefficient of the power-law term (or can even flip its sign), therefore making the enhancement much smaller.

Importantly the 5D small instanton contribution to the axion mass calculated in this paper can 
preserve the Peccei-Quinn solution to the strong CP problem. This is evident because the axion potential arises only from the anomalous coupling $a\, {\rm Tr}[G_{\mu\nu} {\widetilde G}^{\mu\nu}]$, and not from other sources of Peccei-Quinn breaking. In other words, there is no misalignment and the 5D small instanton contributions only scale up the axion potential. Therefore the fact that instantons depend on physics at high energy scales suggests that the axion mass can also be a sensitive probe of UV physics.

\section*{Acknowledgments}
We thank Prateek  Agrawal, Csaba Cs\'aki, Max Ruhdorfer and Misha Shifman for helpful conversations. The work of T.G. is supported in part by the DOE Grant No.DE-SC0011842 at the University of Minnesota, and the Simons Foundation. The research of V.V.K. is partially supported by the STFC consolidated grant ST/T001011/1.  The work of A.P. is supported by the Catalan ICREA Academia Program and  grants FPA2017-88915-P, 2017-SGR-1069 and SEV-2016-0588. Y.S. is supported in part by the NSF grant PHY-1915005. We acknowledge the Munich Institute for Astro- and Particle Physics (MIAPP) of the DFG Excellence Cluster Origins, where this work was initiated. T.G. and A.P. also thank the Kavli Institute of Theoretical Physics in Santa Barbara where part of this work was done. Y.S. also thanks the Aspen Center for Physics for hospitality while this work was in progress.

\appendix
\section{Five-dimensional instanton}
\label{sec:App5D}

Let us consider an $SU(N)$  gauge theory and study the 5D fluctuations $\delta A_M$ around the instanton solution (\ref{eq:5Dinstantonsoln}). We choose a 5D generalization of the 't Hooft gauge~\cite{tHooft:1976snw}:
\begin{equation}
{\cal L}_{GF}=-\frac{1}{2g_5^2}{\rm Tr}\left[D^{(I)}_\mu A_{\mu}+\partial_5  A_5\right]^2\, ,
\end{equation}
where $D^{(I)}_\mu A_\mu=\partial_\mu A_\mu - i  [A^{(I)}_\mu,A_\mu]$, is the gauge covariant derivative evaluated on the instanton background. This requires a 5D Faddeev-Popov ghost term:
\begin{equation}
{\cal L}_{\rm gh}=-\frac{1}{2g_5^2}{\rm Tr}\left[\bar c (-D^{(I)\, 2}_\mu+\partial_5^2)  c\right]\,,
\end{equation}
where $c$ is the ghost field.
Performing a  KK decomposition for the fluctuations of the gauge field $\delta A_\mu$, ghost 
$\delta c$, and 5th component $\delta A_5$:
\begin{eqnarray}
\delta A_\mu(x,y)&=&\delta A_\mu^{(0)}(x)+\sqrt{2} \sum_{n=1}^\infty\delta A_\mu^{(n)}(x)\cos(ny/R)\,,\\
\delta c (x,y)&=&\delta c^{(0)}(x)+\sqrt{2} \sum_{n=1}^\infty\delta c^{(n)}(x)\cos(ny/R)\,,\\
\delta A_5(x,y)&=&\sqrt{2} \sum_{n=1}^\infty\delta A_5^{(n)}(x)\sin(ny/R)\,,
\end{eqnarray}
and replacing $A_\mu\to A_\mu^{(I)}+\delta A_\mu$, 
$A_5\to\delta A_5$ and $c\to\delta c$ in the action (\ref{eq:5Daction}),
we obtain at quadratic order, and after integrating over the extra dimension:
\begin{eqnarray}
S_5&=&S_5^{(I)}-\frac{1}{2g_s^2}\int d^4x \,{\rm Tr}\Big[
\delta A_\mu^{(0)}{\cal M}^{\mu\nu}_A \delta A_\nu^{(0)}+
\delta  \bar c^{(0)}{\cal M}_{\rm gh} \delta c^{(0)}\\
&+&
\sum_{n=1}^\infty\left(\delta A_\mu^{(n)}({\cal M}^{\mu\nu}_A+m_n^2\delta_{\mu\nu}) \delta A_\nu^{(n)}
+\delta  \bar c^{(n)}({\cal M}_{\rm gh}+m_n^2) \delta c^{(n)}
+\delta A_5^{(n)}({\cal M}_5+m_n^2) \delta A_5^{(n)}\right)\Big]\nonumber
\, ,
\end{eqnarray}
where  $g_s$ is defined in (\ref{eq:couplings}), $m_n=n/R$, and the expressions for the gauge boson (gluon) operator ${\cal M}^{\mu\nu}_A$ and ghost operator ${\cal M}_{\rm gh}$ can be found in~\cite{tHooft:1976snw}. 
The  $\delta A_5^{(n)}$ fluctuations behave as 4D scalars of mass $m_n$ in the adjoint representation of the $SU(N)$ group, and therefore ${\cal M}_5$ is the same as the operator expression for a 4D massless scalar ${\cal M}_\Phi$  given in~\cite{tHooft:1976snw}.
 
The existence of $4N$ zero-frequency modes for $\delta A_\mu^{(0)}$, corresponding to those of a 4D instanton
(four associated with  the instanton location ($x_0$), one for its size ($\rho$) and the rest for the orientation in group space), tells us that there are $4N$ eigenstates satisfying  ${\cal M}^{\mu\nu}_A \delta A_\nu^{(0)}=0$. Therefore, 
 since ${\cal M}^{\mu\nu}_A$  is the same for the  zero mode as well as for the $n$-th KK mode, we have that for each KK mode there are $4N$ eigenstates satisfying ${\cal M}^{\mu\nu}_A \delta A_\nu^{(n)}+m^2_n \delta A_\mu^{(n)} =m^2_n\delta A_\mu^{(n)}$, i.e., they have eigenvalues $m^2_n$. Using a 5D Pauli-Villars field of  mass $M\gg 1/R$ to regularize the theory (whose KK squared masses are  $M^2+m_n^2$), we can integrate the path integral over these $4N$  modes and obtain the following contribution to the partition function:
\begin{equation}
\prod_{n=1}^\infty\left(\frac{M^2+m^2_n}{m_n^2}\right)^{2N}\,.
\label{eq:4NKK}
 \end{equation}
The  $4N$ zero-frequency  modes of $\delta A_\mu^{(0)}$ must be treated as collective coordinates, which means including a pre-factor in the integration of the form
\begin{equation}
\int d^4x_0 \int \frac{d\rho}{\rho^5} C[N]\left(\frac{2\pi}{\alpha_s}\right)^{2N} (M\rho)^{4N}\,,
\label{eq:4Nzero}
\end{equation}
where the coefficient $C[N]$ is given by 
\begin{equation}
    C[N] = \frac{C_1\,e^{-C_2 N}}{(N-1)!(N-2)!} ~,
    \label{eq:Cdef}
\end{equation}
and $C_1,C_2$ are order one constants ($C_1=0.466, C_2=1.679$ using Pauli-Villars regularization~\cite{Vainshtein:1981wh}).

In addition to the zero-frequency modes, there are also positive frequency modes for  $\delta A_\mu^{(0)}$ and $\delta A_\mu^{(n)}$ and their corresponding ghosts. In 4D this was calculated in~\cite{tHooft:1976snw}, where it was shown that the massless gauge bosons and ghosts combine to give a contribution equivalent to two real scalars in the adjoint of $SU(N)$. The contribution is approximately
\begin{equation}
 e^{-\frac{N}{3}\ln(M\rho)}\,.
\label{masslessfreqmodes}
 \end{equation}
For the $n$th KK mode we expect a similar contribution where both the transverse part of the gauge bosons and ghosts combine to give the contribution of two real scalars of mass $m_n^2$. In addition this must also be combined with the contribution of the longitudinal component, $\delta A_5^{(n)}$ to provide three real scalars of mass $m_n^2$. Due to their masses, the contribution deviates from the massless result (\ref{masslessfreqmodes}) and the massive contribution was numerically calculated  in Ref.~\cite{Dunne:2005te}. We have used this latter result to obtain the contribution of positive frequency modes of  all $n$th-KK modes:
\begin{equation}
e^{-3 N\sum_{n=1}^\infty\left( \frac{1}{12}\ln\left(\frac{M^2+m^2_n} {m_n^2}\right)+\frac{1}{6}\ln(m_n\rho)+\tilde\Gamma^S_{\rm ren}(m_n\rho)  \right)}=
\prod_{n=1}^\infty\left(\frac{m_n^2}{M^2+m^2_n}\right)^{\frac{N}{4}}
 e^{{N}\xi(R/\rho)\frac{R}{\rho} }\,,
\label{freqmodes}
 \end{equation}
where  $\tilde\Gamma^S_{\rm ren}$ is defined in \cite{Dunne:2005te} and we have used the quite accurate interpolating function given in Eq.~(6.2) of \cite{Dunne:2005te}. The function $\xi(R/\rho)$ defined from (\ref{freqmodes}) is  shown in Fig~\ref{fig:xi}. Notice that for $R\gg \rho$ the function tends to a constant value $\xi(R/\rho) \sim 0.35$, and therefore the exponent  in (\ref{freqmodes}) has a  power-law enhancement $\propto R/\rho$.  
 
Finally, combining the contributions (\ref{eq:4NKK}), (\ref{eq:4Nzero}), (\ref{masslessfreqmodes}) and (\ref{freqmodes}) gives the $SU(N)$ result:
\begin{equation}
\prod_{n=1}^\infty\left(\frac{M^2+m^2_n}{m_n^2}\right)^{\frac{7N}{4}}
\int d^4x_0 \int \frac{d\rho}{\rho^5} C[N]\left(\frac{2\pi}{\alpha_s}\right)^{2N}
 e^{-\frac{2\pi}{\alpha_s}+N\left( \frac{11}{3}\ln(M\rho)
+ \xi(R/\rho)\frac{R}{\rho}\right)}\,.
\label{inst5db}
\end{equation}
To understand the physical implication  of this contribution, it is convenient to write the bare coupling $\alpha_s$  appearing in the exponent of (\ref{inst5db}) as a function of a more physical  gauge coupling. To do so, we first
calculate the one-loop self-energy   of the massless mode $A_\mu^{(0)}$ regularized by  the 5D Pauli-Villars field. 
This  is given by
\begin{equation}
\Pi(q^2)=\frac{1}{g_s^2}-\frac{1}{16\pi^2}\left[
b_0\ln\frac{M^2}{q^2}+
b_{KK}\sum_{n=1}^{\infty}\ln\frac{M^2+m^2_n}{m^2_n}+\Delta(q^2)\right]\,,
\end{equation}
where  $b_0=11N/3$, $b_{KK}=7N/2$.
The value of $\Delta(q^2)$ is independent of $M$ and $m_n$, and for  $q^2\lesssim 1/R^2$  is very small, $\Delta(q^2)\lesssim  0.01$, and can be neglected. We can then define the renormalized one-loop gauge coupling 
of the massless mode $A_\mu^{(0)}$ at the compactification scale as
\begin{equation}
\frac{1}{\alpha_s(1/R)}\equiv 4\pi\Pi(q^2=R^{-2})\simeq \frac{1}{\alpha_s}-\frac{1}{4\pi}\left[
b_0\ln (MR)^2+ b_{KK}\sum_{n=1}^{\infty}\ln\frac{M^2+m^2_n}{m^2_n}\right]\,.
\label{gaugeR}
\end{equation}
Substituting (\ref{gaugeR}) into (\ref{inst5db}), we obtain the $SU(N)$ result 
\begin{equation}
\int d^4x_0 \int \frac{d\rho}{\rho^5} C[N]\left(\frac{2\pi}{\alpha_s}\right)^{2N}
e^{-\frac{2\pi}{\alpha_s(1/R)}+N\xi(R/\rho)\frac{R}{\rho}-b_0\ln\frac{R}{\rho}}\,,
\label{inst5dSUN}
\end{equation}
where we see that the Pauli-Villars mass regulator $M$ has disappeared from the exponent.
The result for $N=3$ is given in (\ref{eq:inst5dSU3}). There still remains the bare coupling $\alpha_s$ in the pre-factor that when written in terms of $\alpha_s(1/R)$ will depend on the regulator mass $M$ (as can be seen from (\ref{gaugeR})). To eliminate this regulator dependence one should go beyond the one-loop level calculation.
As a good estimate for $\rho\lesssim R$, one can replace $\alpha_s$ by $\alpha_s(1/R)$.

\subsection{Fermion contributions}
It is straightforward to incorporate the effect of fermions. For the instanton to contribute to the vacuum energy and the axion potential, fermion zero modes need to be soaked up. Usually this is achieved by lifting zero eigenvalues with insertions of the Higgs vacuum expectation value. In the single Higgs doublet models considered here the instanton contribution can be obtained by soaking up sets of four fermion zero modes with the Higgs propagators (see Figure~\ref{fig:instantondiagram}). As discussed in the main text, this leads to a pre-factor $y_f/(4\pi)$ for each fermion, and (\ref{inst5dSUN}) is now modified by having $b_0=11N/3-2N_f/3$ and   $N\xi(R/\rho) \to  (N-4N_f/3)\xi(R/\rho)$, where $N_f$ is the number of 5D bulk fermions in the fundamental representation. In the case of just boundary fermions, $b_0$ still changes but $N\xi(R/\rho)$ remains unchanged.

\section{4D Deconstruction}
\label{sec:4Ddeconstruction}

In this Appendix we present the calculation of the small instanton contributions using the 4D deconstruction method in \cite{Poppitz:2002ac}. In the notation of \cite{Poppitz:2002ac} the 5D instanton solution (\ref{eq:5Dinstantonsoln}) corresponds to a multi-instanton configuration with winding numbers $(1,1,\dots,1)$. Using this solution, the effective action $S_{\rm eff}$ is given by the $SU(N)$ generalization of equation (27) in \cite{Poppitz:2002ac}, namely\footnote{Note the coefficient of the log term in Eq.(27) of \cite{Poppitz:2002ac} should be 8 corresponding to the number of bosonic zero modes for $SU(2)$.}
\begin{equation}
    S_{{\rm eff}}(R\gg \rho\gg 1/\Lambda_5)= \frac{2\pi}{\alpha_s(1/R)} 
    -\xi_0 N\frac{R}{\rho} +  \xi_1 4 N \ln \frac{R}{\rho}~,
    \label{eq:Seff}
\end{equation}
where $\Lambda_5$ is the 5D cutoff. In the step approximation, the numerical factor $\xi_0=\xi_1=1$ for a pure 5D gauge theory compactified on a circle of radius $R$. The coefficient of the $\ln$ term, $4N$ is the number of bosonic instanton zero modes.

In orbifold compactifications $A_5^{(0)}$ is projected out and there are half as many KK states, therefore the expressions of \cite{Poppitz:2002ac} must be modified. We find $\xi_0=\frac{1}{2}$ and $\xi_1=\frac{47}{48}$. The step approximation can be improved by including threshold corrections which can further modify $\xi_{0,1}$ (an example with a bulk scalar in the fundamental representation shown in Section 5 of \cite{Poppitz:2002ac} reveals an approximately 30\% change). The expression (\ref{eq:Seff}) assumes that $1/\rho$ is sufficiently below the top of the KK tower, and also that $R/\rho \geq {\cal O}(10)$ so there is a large number of KK modes lighter than $1/\rho$ that feel the instanton. Substituting (\ref{eq:Seff}) into the partition function gives the leading behaviour
\begin{equation}
        e^{-S_{\rm eff}} = e^{-\frac{2\pi}{\alpha_s(1/R)} + \xi_0 N \frac{R}{\rho}-\xi_1 4 N \ln\frac{R}{\rho}}~.
        \label{eq:expSeff}
\end{equation}
This expression reveals that there is an exponential enhancement of the instanton density for small
instantons of size $\rho < R$ (assuming $\xi_0 >0$) due to the  power-law term in (\ref{eq:Seff}).

It is instructive to compare this effect with the 4D moose models of \cite{Agrawal:2017ksf,Agrawal:2017evu}.
In that model the instanton density at each site is suppressed by a factor $e^{-2\pi/\alpha_1} = e^{-2\pi/(N \alpha_s)}$, since $1/\alpha_s = 1/\alpha_1 + 1/\alpha_2 + \dots \approx N/\alpha_1$, assuming approximately equal site gauge couplings and $N$ sites. Clearly as $N\rightarrow \infty$ the large exponential suppression is mitigated. Alternatively this is equivalent to the freedom in choosing a large site gauge coupling $\alpha_i$.  Instead in the extra dimension a single axion receives contributions from the $N$ sites which sum to give the factor $e^{-2\pi/\alpha_s}$. There is no factor of $N$ in the denominator of the exponent as found in the 4D moose model, and instead a sum over KK modes gives the power-law enhancement in (\ref{eq:Seff}).

To obtain the full instanton contribution to the partition function, the instanton density must be integrated over the instanton size $\rho$. The integration over the instanton size is divergent but is regulated by the finite size of the extra dimension. Assuming an orbifold compactification ($\xi_0=\frac{1}{2}, \xi_1=\frac{47}{48}$) we obtain
\begin{eqnarray}
    &&  \int_{1/\Lambda_5}^R \frac{d\rho}{\rho^5}~C[3] \left(\frac{2\pi}{\alpha_s(1/\rho)}\right)^6 
      e^{-\frac{2\pi}{\alpha_s(1/R)} 
      + \frac{3}{2}\frac{R}{\rho} -(b_0+\frac{3}{4}) \ln \frac{R}{\rho}}\nonumber\\
  &&\qquad \approx C[3] \left(\frac{2\pi}{\alpha_s(1/R)}\right)^6 \frac{2}{3}\frac{e^{-\frac{2\pi}{\alpha_s(1/R)}+\frac{3}{2}\Lambda_5 R}} {(\Lambda_5 R)^{b_0-9/4}} \frac{1}{R^4}\,,
       \label{eq:rhointegral}
\end{eqnarray}
where $C[3]\simeq 1.5\times 10^{-3}$ using (\ref{eq:Cdef}), $b_0 = 11$, and $\Lambda_{IR}$ is defined in (\ref{eq:LamIR}).  This expression is consistent with the 5D calculation (\ref{eq:Kconstant}). If the step approximation is improved then the exponent in $e^{\frac{3}{2}\Lambda_5 R}$ is reduced by approximately $30\%$ to become $\sim e^{\Lambda_5 R}$.

Note that the running of the gauge coupling with instanton size, $\alpha_s(1/\rho)$ has only been crudely approximated in (\ref{eq:rhointegral}). The actual running coupling is given by:
\begin{eqnarray}
      \frac{2\pi}{\alpha_s(1/\rho)} &=&  \frac{2\pi}{\alpha_s(1/R)}  + b_0 \ln \frac{R}{\rho} - b_{KK} \sum_{n=1}^{K(\rho)} \ln \rho m_n~,\nonumber\\
      &=& \frac{2\pi}{\alpha_s(1/R)}  + b_{KK}  \frac{R}{\rho} + \left(b_0- \frac{1}{2} b_{KK}\right) \ln \frac{R}{\rho}~,
      \label{eq:alpharunning}
\end{eqnarray}
where $K(\rho)$ is the number of KK levels lighter than $1/\rho$, and $b_{KK}=\frac{7}{2} N$ is the KK contribution to the $\beta$-function arising from a massive gauge boson~\cite{Dienes:1998vh}. In the second line
of (\ref{eq:alpharunning}) we have used the fact that in flat space $m_n = n/R$ and $\sum_{n=1}^{K(\rho)} \ln \rho m_n = -\frac{R}{\rho} -\frac{1}{2}\ln \frac{\rho}{R}$. Using (\ref{eq:alpharunning}) 
clearly makes the integrand in (\ref{eq:rhointegral}) larger for $1/\Lambda_5 \leq \rho \leq R$ and therefore 
the analytic expression in (\ref{eq:rhointegral}) is a conservative lower limit. Thus to obtain a sizeable contribution to the integral in (\ref{eq:rhointegral}), the exponential suppression $e^{-\frac{2\pi}{\alpha_s(1/R)}}$ must be overcome by having a large number of KK states, $\Lambda_5 R$.

\subsection{Fermion contributions}
Next we consider fermions propagating in the 5th dimension compactified on an orbifold. For each chiral SM field we need to introduce a Dirac fermion in the 5D bulk. Thus assuming $N_f$ quark flavors there are $2N_f$ Dirac fermions at each KK level.  Using the results in \cite{Poppitz:2002ac} the effective action for $SU(N)$ becomes:
\begin{equation}
    S_{\rm eff}(R\gg \rho\gg 1/\Lambda_5)= \frac{2\pi}{\alpha_s(1/R)} 
    - \left(\frac{1}{2}N-\frac{2}{3} N_f\right) \frac{R}{\rho} +  \left(b_0+\frac{1}{4} N -\frac{1}{3} N_f \right)\ln \frac{R}{\rho}~,
    \label{eq:Seff_fermion}
\end{equation}
where $b_0=7$ is the QCD contribution from the gauge boson and fermion zero modes.
Compared to the pure YM case given in (\ref{eq:Seff}) the enhancement in the exponential 
factor $e^{-S_{\rm eff}}$  from the power-law $R/\rho$ term is now reduced as the number of flavors 
increases. In fact for $N=3$ there is only an enhancement for $N_f\leq 2$.

The fermion zero modes again lead to a suppression in the instanton vacuum diagrams due to Yukawa couplings and Higgs loops (assuming the Higgs is confined to the boundary). Using (\ref{eq:rhointegral}) and the suppression factor in (\ref{eq:kappaf}) the axion mass ratio in the case of bulk fermions then becomes:
\begin{equation}
     \frac{m_a}{m_{a,QCD}} \simeq \frac{\sqrt{2\kappa_f C[3]}}{\sqrt{\frac{3}{2}-\frac{2}{3}N_f}}
     \left(\frac{2\pi}{\alpha_s(1/R)}\right)^3 \frac{(m_u+m_d)}{\sqrt{m_u m_d}}\frac{1}{m_\pi f_\pi R^2}\frac{e^{-\frac{1}{2}\left(\frac{2\pi}{\alpha_s(1/R)}-\frac{3}{2}+\frac{2N_f}{3}\right) \Lambda_5 R}}{(\Lambda_5 R)^{19/8-N_f/6}}\,,
     \label{eq:bulkfermionratio}
\end{equation}
where (\ref{eq:axionfermionmass}) has been used and $N_f$ in this expression refers to the number of bulk fermion flavors in the fundamental representation. In particular when $N_f=0$ there are only boundary fermions (with $b_0=7$) and we are consistent with the relation (\ref{eq:boundaryfermionratio}). Recall that for the 4D deconstruction we have used the step approximation, and corrections to this method will reduce the pure YM exponent from $3/2$ to 
$\sim 1$.

\bibliographystyle{JHEP}
\bibliography{references}

\end{document}